%% file: amia.tex
\documentclass{amia}
\usepackage{lipsum} 
\usepackage{graphicx}
\usepackage{booktabs}
\usepackage{multirow}
\usepackage{colortbl}
\usepackage{xcolor}
\usepackage{algorithm}
\usepackage{algorithmic}
\usepackage{amssymb}
\usepackage{amsmath}
\usepackage{tabularx}
\usepackage{paralist}
\usepackage{soul}
\usepackage[T1]{fontenc}
\usepackage{wrapfig}

\begin{document}

\title{Higher-order Interaction Matters: Dynamic Hypergraph Neural Networks for Epidemic Modeling}

\author{Songyuan Liu\textsuperscript{1},  Shengbo Gong\textsuperscript{1},  Tianning Feng\textsuperscript{1},\\
Zewen Liu\textsuperscript{1}, Max S.Y. Lau, PhD\textsuperscript{2}, Wei Jin, PhD\textsuperscript{1}}
\institutes{
    \textsuperscript{1}Department of Computer Science, Emory University, Atlanta, GA\\
    \textsuperscript{2}Department of Biostatistics and
    Bioinformatics, Emory University, Atlanta, GA
}

\maketitle
\section*{Abstract}
\vspace{-4mm}
\textit{The ongoing need for effective epidemic modeling has driven advancements in capturing the complex dynamics of infectious diseases. Traditional models, such as Susceptible-Infected-Recovered, and graph-based approaches often fail to account for higher-order interactions and the nuanced structure pattern inherent in human contact networks. This study introduces a novel Human Contact-Tracing Hypergraph Neural Network framework tailored for epidemic modeling called EpiDHGNN, leveraging the capabilities of hypergraphs to model intricate, higher-order relationships from both location and individual level. Both real-world and synthetic epidemic data are used to train and evaluate the model. Results demonstrate that EpiDHGNN consistently outperforms baseline models across various epidemic modeling tasks, such as source detection and forecast, by effectively capturing the higher-order interactions and preserving the complex structure of human interactions. This work underscores the potential of representing human contact data as hypergraphs and employing hypergraph-based methods to improve epidemic modeling, providing reliable insights for public health decision-making.}

\input{section/intro}
\input{section/related}
\input{section/formulation}
\input{section/method}

\input{section/experiment}

\input{section/conclusion}

\makeatletter
\renewcommand{\@biblabel}[1]{\hfill #1.}
\makeatother

\bibliographystyle{vancouver}

\input{amia.bbl}
\end{document}

%% file: section/intro.tex
\section{Introduction}
\vspace{-3mm}
Since the onset of the COVID-19 pandemic, there has been a growing interest in studying epidemiological models\cite{rodríguez2022datacentricepidemicforecastingsurvey, liu2024review, 9387581}. Understanding and managing infection outbreaks is crucial for public health. Traditional mechanistic models like Susceptible-Infected-Recovered (SIR), which mathematically describe the transmission mechanisms of infectious diseases, often suffer from limitations of oversimplified or fixed assumptions, leading to sub-optimal predictive power and inefficiency in capturing complex epidemic patterns~\cite{song2020reinforced,deng2019graphmessagepassingcrosslocation}.

Motivated by these limitations, sequential models such as GRU~\cite{chung2014empirical} and LSTM~\cite{yu2019review} are used to model temporal relations in a data-driven manner. Compared to mechanistic models, sequential models have demonstrated superior performance in forecasting infection counts~\cite{rodríguez2022datacentricepidemicforecastingsurvey,rodriguez2023einns,liu2025capecovariateadjustedpretrainingepidemic}. However, these models often struggle to incorporate spatial dependencies, such as human mobility patterns and geographical distributions, which play a crucial role in epidemiology modeling~\cite{liu2024review}. Mobility data captures how individuals move and interact across different locations, influencing disease transmission dynamics beyond simple temporal trends. To enhance the ability to capture both spatial and temporal information, graph-based approaches have emerged as a popular tool in epidemic research. Graph Neural Networks (GNNs)~\cite{yu2018spatio, liu2023attentionbasedspatialtemporalgraphconvolutional} have become popular for their ability to model human mobility patterns. They achieve this by representing nodes as regions and weighted edges as mobility volume, effectively capturing movement between locations. Through a message-passing mechanism, GNNs enable nodes to share information with their neighbors, allowing for a more comprehensive understanding of mobility patterns. Additionally, by leveraging dynamic graph modeling and dynamic GNNs, they can further represent changes in human movement over time, enhancing their ability to model relational dynamics within mobility networks.~\cite{zheng2024epidemiologyinformedgraphneuralnetwork, Qiu_2024, deng2019graphmessagepassingcrosslocation}

\begin{figure}[t!bp!]
\hspace{-1pt}
\centering
\includegraphics[width=475pt]{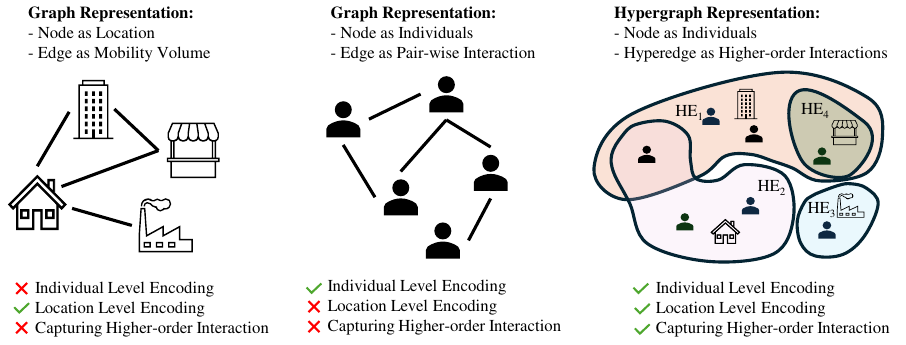}
\vspace{-2.2em}
\caption{Illustration of how various graph encoding methods can be employed to capture complex interactions. Hypergraphs, in particular, offer significant advantages over traditional graphs by retaining both individual-level and location-level information, while also capturing higher-order interactions. This enriched representation facilitates a more nuanced understanding of epidemic dynamics.}
\label{fig:Hypergraph_illustration}
\vspace{-0.2cm}
\end{figure}

Despite the utility of GNN-based methods, they primarily focus on pair-wise interactions and therefore neglect the higher-order interactions that are inherent in actual human contact networks~\cite{Tan_2023, feng2019hypergraph, h2abm}. Specifically, higher-order interactions refer to interactions or contacts that involve more than two individuals simultaneously in the context of epidemic modeling~\cite{kim2024survey}. For example, public transportation, workplaces, and schools are shared spaces where groups of people interact following higher-order transmission dynamics. As illustrated in Figure~\ref{fig:Hypergraph_illustration}, while standard graphs can model these interactions by representing individuals as nodes and forming fully connected subgraphs for each group, this approach is often inefficient and obscures the true higher-order structure. In contrast, hypergraphs provide a more natural and explicit way to represent higher-order interactions through hyperedges, eliminating the artificial clique. Additionally, hypergraphs can model overlapping interactions by representing locations as hyperedges, encompassing multiple individuals simultaneously. These enhancements can lead to more accurate and interpretable modeling of epidemics than standard graphs~\cite{Tan_2023,feng2019hypergraph, 10.1007/978-3-031-26422-1_28}.

As illustrated in Figure~\ref{fig:Hypergraph_illustration}, a fundamental limitation of prior graph-based approaches lies in their inability to simultaneously preserve both individual-level and location-level information, as well as their failure to capture higher-order interactions. These shortcomings significantly hinder the accurate modeling of real-world human contact patterns, which are essential for understanding and predicting the spread of infectious diseases. To address this, we propose \textbf{EpiDHGNN}, a novel framework that models human contact data as dynamic hypergraphs. This approach enables the encoding of complex, higher-order interactions and supports a richer, more granular representation of epidemic dynamics. The major contributions of this paper are threefold:

\begin{compactenum}[(1)]
    \item  We propose a novel method to model human contact as dynamic hypergraphs, which encodes nodes as individuals and hyperedges as locations, leveraging both granular level information and higher-order interactions. 
    \item We develop EpiDHGNN, a model tailor-made for epidemic modeling with a self-supervised contact-pattern awareness module, capturing the higher-order interactions and contact patterns that are inherent in human contacts.
    \item Extensive experiments are conducted to demonstrate the superiority of encoding human contact as hypergraphs, as well as the effectiveness of our proposed models in various epidemic tasks. 
\end{compactenum}

%% file: section/related.tex
\section{Related Work}
\vspace{-3mm}
\textbf{Mechanistic Epidemic Modeling}: In the past, when data was not sufficiently recorded, scientists were unable to build empirical models that successfully captured the dynamics of epidemics. Empirical models rely heavily on accurate and comprehensive data to make predictions and understand patterns. In contrast, mechanistic models\cite{Lessler2016Mechanistic, D_Agostino_McGowan_2021} are designed to capture the underlying complexity of infections and the recovery processes of various diseases, even with limited data. Among these, the compartmental model, exemplified by the Susceptible-Infectious-Recovered (SIR) model\cite{hethcote2000mathematics}, is considered one of the most popular and successful. The model divides the population into compartments based on their disease status, allowing for a structured and systematic analysis of epidemic progression. It uses two parameters, $\beta$ and $\gamma$, to account for the infection and recovery rates respectively. The model' ability to incorporate epidemiological principles makes it particularly valuable for understanding and predicting the course of infectious diseases. 

\textbf{Graphs for Epidemic Modeling}: Recent advancements in dynamic graph modeling have underscored the utility of such models in epidemic source detection and spread prediction. Initially developed for traffic forecasting, dynamic graph models have been rapidly adapted to epidemiological contexts, where nodes represent geographical locations.\cite{liu2024review,deng2019graphmessagepassingcrosslocation,xie2022epignnexploringspatialtransmission,wan2024epidemiology} However, these methods overlook the higher-order interaction inherent in human contact networks. Furthermore, previous works also focus on utilizing the dynamic message-passing (DMP) inference algorithm and network centrality as a tool for statistical inference to estimate the origin of an epidemic outbreak.\cite{Lokhov2014DMP, Yu_2022,Tan_2023} Such algorithms iteratively transmit messages along network edges, updating each node's state probabilities based on the states of its neighbors. However, they operate on static graphs, thereby overlooking the inherent dynamics of contact networks in human societies.


\textbf{Hypergraphs for Epidemic Modeling}: Similar to graphs, hypergraphs can be utilized in epidemic modeling while encompassing higher-order interactions. \cite{abc,PhysRevResearch.2.023032,bodó2015sis} In the pathogen propagation function proposed by Hypergraph-HeterSIS\cite{h2abm}, the infection state of each node is first aggregated to hyperedge, which is then followed by a nonlinear function $f$ to remove linearity. The result is then mapped back to node level to provide the next step update. The method has been shown that hypergraph-based approaches are better at capturing the structural differences in contact networks and improving the accuracy of infection dynamics modeling. However, these approaches are based on variable calibration, therefore neglecting the higher-level representation generated through deep learning approaches.\cite{feng2019hypergraph,chien2022allset,agarwal2022think}

%% file: section/formulation.tex
\section{Formulation}
\vspace{-3mm}
\textbf{3\hspace{0.3mm}.\hspace{0.3mm}1 Hypergrpah}:
A hypergraph is a higher-order representation of a graph where an edge can connect any number of vertices. Formally, a hypergraph \( G = (V, \mathcal{E}, \mathbf{X}) \) consists of a set of nodes \( V \), a set of hyperedges \( \mathcal{E} \), where each hyperedge is a subset of \( V \), and a feature matrix $\mathbf{X} \in \mathbb{R}^{|V| \times d}$, where each row encodes the node feature. The hypergraph structure can be described by an incidence matrix \( \mathbf{H} \in \mathbb{R}^{|\mathcal{E}| \times N} \), where \( \mathbf{H}_{i,j} = 1 \) only when the node \( v_i \) is incident to the edge \( e_j \).

\textbf{3\hspace{0.3mm}.\hspace{0.3mm}2 Dynamic Hypergrpah}: A dynamic hypergraph is an extension of a hypergraph that evolves over time, consisting of a sequence of hypergraphs observed over \( T \) discrete time stam\text{ps}. Formally, a dynamic hypergraph is represented as $G^{(0:T)} = \{ G^{(0)}, G^{(1)}, \dots, G^{(T)} \}$, where each hypergraph \( G^{(t)} = (V^{(t)}, \mathcal{E}^{(t)}, \mathbf{X}^{(t)})\) denotes the hypergraph at time stamp \( t \in [0:T]\). Here, \( V^{(t)} \) is the set of nodes, \( \mathcal{E}^{(t)} \) is the set of hyperedges, and $\mathbf{X}^{(t)}$ denotes the node features at time \( t \). It is worth noting that \textbf{both} the graph structure and node features are dynamic, since in some wor\text{ks}, dynamic graphs have static features.

\textbf{3\hspace{0.3mm}.\hspace{0.3mm}3 Epidemic Tas\text{ks}}:
Consider an input of a dynamic hypergraph \( G^{(0:T)} = \{ G^{(0)}, \dots, G^{(T)} \} \), where each node represents an individual and each hyperedge represents a location. At an arbitrary time stamp \( t \), the nodes in a hyperedge \( e^{(t)} \in \mathcal{E}^{(t)} \) represent a single contact between these entities. Each hypergraph \( G^{(t)} \) is associated with an individual state matrix \( \mathbf{X}^{(t)} \in \mathbb{R}^{N \times d} \), where $d$ is the feature dimension of the individual. For example, in the SIR setting, $d$ can consist of three dimensions, which correspond to the \(\{\text{Suspected}, \text{Infected}, \text{Recovered}\}\) status of a specific individual. 

Additionally, we define three time stam\text{ps} to clarify the time interval of our downstream epidemic tas\text{ks}. $[0:\text{tsh}]$ where $\text{tsh}$ stands for Time Stamp Hidden; $[\text{tsh}:\text{ks}]$ where $\text{ks}$ stands for Known Time Stamp; $[\text{ks}:\text{ps}]$ where $\text{ps}$ stands for Prediction Time Stamp. The three time stam\text{ps} are ordered such that $0 \leq \text{tsh} \leq \text{ks} \leq \text{ps} \leq T$. Note that for a time stamp $t \in T$, when $t < \text{tsh}$, only contact hypergraph can be observed. When $\text{tsh} \leq t \leq \text{ks}$, both contact hypergraph and individual state can be observed. When $t > \text{ps}$, neither contact hypergraph nor individual state can be observed. An illustration of the three time stam\text{ps} is shown in Figure~\ref{fig:Model Architecture}.

\noindent\textbf{3\hspace{0.3mm}.\hspace{0.3mm}3\hspace{0.3mm}.\hspace{0.3mm}1 Source Detection}: The source detection task focuses on identifying the initial node responsible for the spread of an epidemic, often referred to as "patient zero." Given the dynamic hypergraph \( G^{(0:\text{ks})} \), or its incidence matrix $\mathbf{H}^{(0:\text{ks})}$, and the corresponding state matrix \( \mathbf{X}^{(\text{tsh}:\text{ks})} \), we aim to infer the likelihood distribution over all nodes at the initial time stamp \( T=0 \) using a model $f$ parametrized by weight $\theta$. Mathematically, we are interested in using $f_\theta$ to estimate the distribution:
\[
f_\theta(\mathbf{H}^{(0:\text{ks})}, \mathbf{X}^{(\text{tsh}:\text{ks})}) \approx p(\mathbf{X}^{(0)} | \mathbf{H}^{(0:\text{ks})}, \mathbf{X}^{(\text{tsh}:\text{ks})}).
\]
This task leverages both the structural properties of the hypergraph and the temporal evolution of the feature ma\text{ps} to backtrack the probable origin of the epidemic. The node labels $y_{\text{detect}}$ are extracted from specific columns of $\mathbf{X}^{(0)}$ that represent the infection state — for example, \textbf{the "infected" column} in the case of the SIR model. To optimize the model, we'll use weighted binary cross-entropy loss between the predictions and node labels $y_{\text{detect}}$, where $w_1 = \frac{|V|}{|y_{\text{detect}} = 1|}$ and $w_0 = \frac{|V|}{|y_{\text{detect}} = 0|}$
\[
\mathcal{L}_\text{detect}(\theta) = - \frac{1}{|V|} \sum_{v \in V} \left[ w_1 y_{\text{detect}} \log(f_\theta) + w_0 (1 - y_{\text{detect}}) \log(1 - f_\theta \right)],
\]
\vskip 0.2em
\noindent \textbf{3\hspace{0.3mm}.\hspace{0.3mm}3\hspace{0.3mm}.\hspace{0.3mm}2 Infection Forecasting}: Forecasting tas\text{ks} in epidemics are usually defined as finding the total number of infections and recoveries in a range of future time stam\text{ps}. This is because previous approaches encode nodes as areas, neglecting the individual level information. On the other hand, when using human contact hypergraphs, we can deduce a more fine-grained forecasting on an individual level. Therefore, we treat our forecast task as a binary node classification task. The node labels are defined as whether a node is in the "infected" state. Using the SIR model as an example, each node can be in one of three states, and only when it is in the "infected" state does the label become True. Here we are interested in using a model $g$ parametrized by $\theta$ to estimate the distribution:
\[
g_\theta(\mathbf{H}^{(0:\text{ks})}, \mathbf{X}^{(\text{tsh}:\text{ks})}) \approx p(\mathbf{X}^{(\text{ks}+1, \text{ps})} | \mathbf{H}^{(0:\text{ks})}, \mathbf{X}^{(\text{tsh}:\text{ks})})
\]
Similar to source detection, we'll use the binary cross-entropy loss between the predictions and node labels $y_{\text{forecast}}$. The labels are extracted from specific columns of $\mathbf{X}^{(\text{ks} + 1:\text{ps})}$ that represent the infection state, similar to source detection label extraction.
\[
\mathcal{L}_{\text{forecast}}(\theta) = - \frac{1}{|V|} \sum_{v \in V} \left[ y_{\text{forecast}} \log(g_\theta) + (1 - y_{\text{forecast}}) \log(1 - g_\theta \right)],
\]

%% file: section/method.tex
\section{Method}
\vspace{-3mm}
In this section, we will formulate our proposed model EpiDHGNN, which serves as $f_\theta$ and $g_\theta$ defined in Section 3.3. Here we define $t \in T_{\text{interest}}$ where  $T_{\text{interest}}$ is the corresponding input interval for $\mathbf{H}$ defined in Sections 3.3.1 and 3.3.2.
\begin{figure}[tbp!]
\hspace{-1pt}
\centering
\includegraphics[width=475pt]{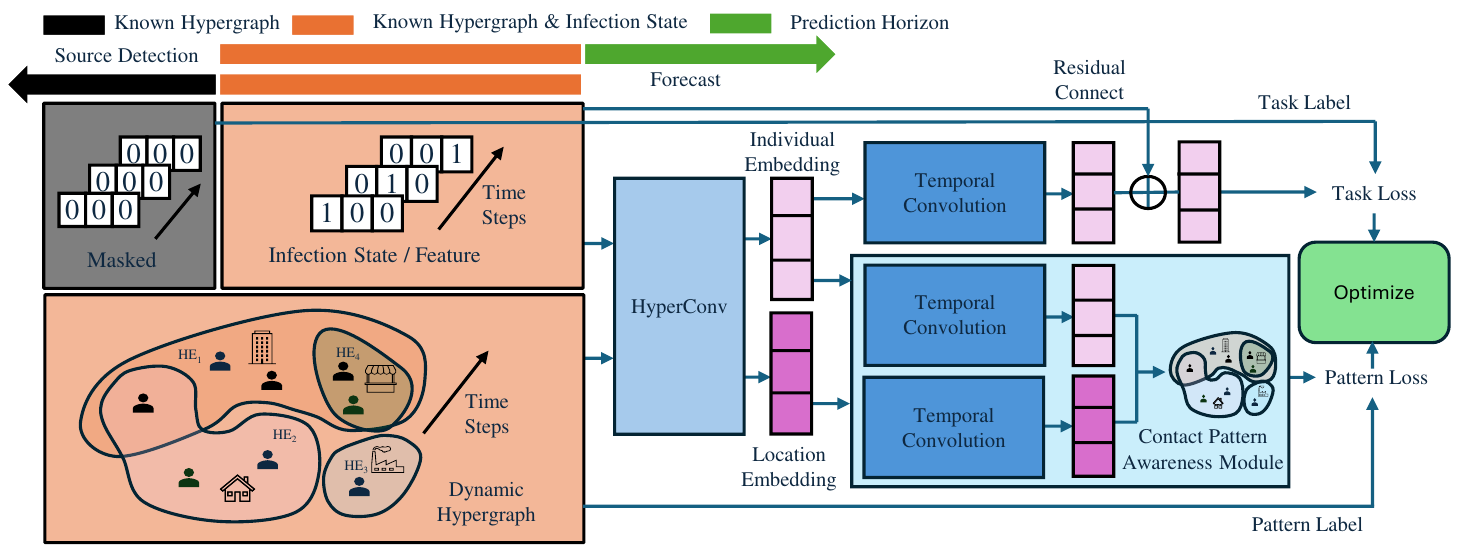}
\vspace{-0.6cm}
\caption{Model Architecture of proposed EpiDHGNN model. The arrows in the top left corner refers to the three time stamps defined in Section 3.3, where $[0:tsh]$ is the black interval, $[tsh:ks]$ is the orange interval, and $[ks+1:ps]$ is the green interval. All individual state is masked to 0 in $[0:tsh]$ as shown in the top left black module. Corresponding inputs for source detection and forecast defined in Section 3.3.1 and 3.3.2 is then feed to the model as input. The light blue HyperConv module in defined in Section 4.1; the dark blue temporal convolution module is defined in Section 4.1; and the contact pattern awareness module is defined in Section 4.2.}
\label{fig:Model Architecture}
\end{figure}

\textbf{EpiDHGNN} is a novel dynamic hypergraph neural network that models epidemic dynamics by jointly capturing spatial, temporal, and structural contact patterns. As illustrated in Figure~\ref{fig:Model Architecture}, EpiDHGNN follows an encoder-decoder paradigm and integrates both supervised and self-supervised learning objectives to enhance the quality of node embeddings. The framework is composed of three key components: (1) a \textit{Spatial-Temporal Encoder} based on hypergraph and temporal convolution that aggregates high-order relational information among individuals and  captures progression patterns over time, (2) a novel self-supervised \textit{Contact Pattern Awareness Module} that encourages structural consistency by reconstructing dynamic contact patterns. This architecture allows the model to learn robust spatiotemporal embeddings of infection states, improving performance on downstream tasks such as epidemic forecasting and source detection.

\textbf{4\hspace{0.3mm}.\hspace{0.3mm}1 Spatio-Temporal Encoding}:  
To capture spatial dependencies at each time step, we adopt the hypergraph convolution operator proposed in HGNN~\cite{feng2019hypergraph}, which aggregates information from high-order neighbors via node-edge-node message passing. Formally, a single hypergraph convolutional layer consists of:
\begin{equation}
    \mathbf{X}_{\text{edge}}^{l+1,t} = \mathbf{D}_e^{-1} \mathbf{H}^{T,t} \mathbf{D}_v^{-1/2} \mathbf{X}_{\text{node}}^{l,t} \Theta,
\end{equation}
\begin{equation}
    \mathbf{X}_{\text{node}}^{l+1,t} = \mathbf{D}_v^{-1/2} \mathbf{H}^t \mathbf{W} \mathbf{D}_e^{-1} \mathbf{H}^{T,t} \mathbf{D}_v^{-1/2} \mathbf{X}_{\text{node}}^{l,t} \Theta,
\end{equation}
where $\mathbf{H}^t$ is the incidence matrix at time $t$, $\mathbf{D}_v$ and $\mathbf{D}_e$ are diagonal node and edge degree matrices, $\mathbf{W}$ is an optional hyperedge weighting matrix, and $\Theta$ denotes learnable parameters. By stacking $L$ such layers, the node embeddings can incorporate information from $L$-hop neighborhoods. After applying the convolution over all time steps, the node and edge embeddings are concatenated along the temporal axis:
\begin{equation}
    \mathcal{X}_{\text{node}}^{L} = [\mathbf{X}_{\text{node}}^{L,t_0} \mid \mathbf{X}_{\text{node}}^{L,t_1} \mid \ldots], \quad 
    \mathcal{X}_{\text{edge}}^{L} = [\mathbf{X}_{\text{edge}}^{L,t_0} \mid \mathbf{X}_{\text{edge}}^{L,t_1} \mid \ldots], \quad 
    \forall t \in T_{\text{interest}}.
\end{equation}

To capture temporal dependencies, we apply standard 1D convolution across the temporal axis, similar to the temporal convolutional networks introduced in \cite{lea2016temporalconvolutionalnetworksunified} and later adapted for dynamic graphs in \cite{guo2019attention}. Letting time correspond to the width dimension, individuals to the height, and feature channels to the depth, the operation is defined as:
\begin{equation}
    \hat{\mathbf{X}}_{\text{node}} = \sigma \left( \Phi_{\text{temporal}}^{k} \circledast \mathcal{X}_{\text{node}}^{L} \right),
\end{equation}
where $\circledast$ denotes convolution, $\Phi_{\text{temporal}}^k$ is the convolution kernel of width $k$, and $\sigma$ is a non-linear activation function. This enables the model to learn progression patterns across time.

\textbf{4\hspace{0.3mm}.\hspace{0.3mm}2 Self-supervised Contact Pattern Awareness Module}:  
In a societal setting, human interactions occur with varying probabilities based on social structures and daily routines. For instance, individuals are highly likely to engage in frequent interactions with family members or colleagues at home or in the workplace, while social encounters with friends or individuals sharing similar interests may occur less frequently, such as on a weekly or monthly basis in clubs or shopping centers. To leverage this information, we propose a self-supervised \textit{Contact Pattern Awareness Module}, designed to predict human interactions within the epidemic framework. Given a sequence of $k$ hypergraphs $\{G_t\}_{t=t_0}^{t_0+k-1}$ starting from a randomly selected time step $t_0$, the module aims to reconstruct the hypergraph at the final time step, $G_{t_0+k}$, using information from the preceding $k-1$ hypergraphs. Successful reconstruction enables HGNN to inject human-contact-aware inductive bias into the learned node embeddings $\mathcal{X}_{\text{node}}^{L}$, thereby improving performance in subsequent epidemic forecasting tasks.

To effectively capture these patterns, we utilize both the individual embeddings and location embeddings obtained from Section~4.1. These embeddings are then processed using the temporal convolution framework introduced in Section~4.2, enabling the model to extract temporal dependencies. Finally, the refined embeddings are passed through a fully connected layer to produce a confidence score for contact prediction. Mathematically, this operation is formulated as:
\begin{equation}
    \mathbf{s} = \sigma \left( \text{MLP}\left( \Phi_{\text{pattern}}^k \circledast \mathcal{X}_{\text{node}}^{L} * \Phi_{\text{pattern}}^k \circledast \mathcal{X}_{\text{edge}}^{L} \right) \right),
\end{equation}
where $*$ denotes element-wise multiplication and $\sigma$ is a sigmoid activation. The output $\mathbf{s}$ is a predicted score for individual-location contact. The module is trained with binary cross-entropy loss on a balanced set of positive (observed) and negative (random) samples:
\begin{equation}
    \mathcal{L}_{\text{pattern}} = - \frac{1}{N} \sum_{i=1}^{N} \left[ y_i \log s_i + (1 - y_i) \log (1 - s_i) \right],
\end{equation}
where $y_i \in \{0, 1\}$ is the ground-truth label indicating whether contact occurred.

\textbf{4\hspace{0.3mm}.\hspace{0.3mm}3 EpiDHGNN}: Given an input sequence of dynamic hypergraphs $\{\mathbf{H}^t\}_{t=t_0}^{t_0+ks}$ and the corresponding individual features $\{\mathbf{X}^{t}\}_{t=t_0}^{t_0+ks}$, EpiDHGNN first encodes node representations using the spatial and temporal modules (Sections~4.1), and then incorporates initial state residuals before calculating task-specific loss. The overall loss is a combination of task-specific supervision and the self-supervised contact pattern loss:
\begin{equation}
    \mathcal{L} = \alpha ~\mathcal{L}_{\text{task}} + (1 - \alpha) \mathcal{L}_{\text{pattern}},
\end{equation}
where $\alpha$ is a weighting hyperparameter that balances predictive accuracy and structural awareness.

%% file: section/experiment.tex
\section{Experiment}
\vspace{-3mm}
In this section, we perform analysis on the datasets and conduct experiments to evaluate the proposed model. We will focus on the following research questions:

\begin{compactenum}[\textbullet]
    \item \textbf{RQ1}: Does EpiDHGNN outperforms baseline dynamic graph models in various epidemic tasks?
    \item \textbf{RQ2}: Does the contact pattern awareness module facilitate the overall performance of EpiDHGNN?
    \item \textbf{RQ3}: Is contact patterns successfully captured? To what aspect of the task does the module helps the most?
    \item \textbf{RQ4}: Beyond individual-level prediction, can EpiDHGNN capture population-level infection dynamics over time?
\end{compactenum}

\begin{wraptable}{R}{0.25\linewidth}
    \vspace{-0.5em}
    \centering
    \caption{Dataset Summary}
        \vspace{-0.8em}
    \label{tab:dataset_summary}
    \resizebox{\linewidth}{!}{%
        \begin{tabular}{lcc}
            \toprule
            \textbf{Metric} & \textbf{UVA} & \textbf{EpiSim} \\
            \midrule
            Individuals & 2,500 & 10,000 \\
            Locations & 500 & 11 \\
            Time Steps & 169 & 47 \\
            Contacts & 94,134 & 664,177 \\
            \bottomrule
        \end{tabular}%
    }
\end{wraptable}
\textbf{Data Description:}
We assess the performance of baseline models and our proposed model on both graphs and hypergraphs settings. Because of the privacy nature of human contact data, we used both real-world and synthetic data. The University of Virginia \textbf{UVA} dataset includes an extensive collection of clinical metadata sourced from the Epic-based SQL database at the UVA hospital. The interactions are derived from Electronic Healthcare Records (EHRs), which document the timing and locations of encounters between patients and healthcare workers (HCWs). We utilized the real-world infection case calibrated pathogen parameters provided by Anand etc. \cite{h2abm} to retrieve the patient infection states through simulations.  The \textbf{EpiSim} dataset is based on the Mobility Intervention of Epidemic's Simulator\cite{simulation_tool}, which models human movement and disease transmission. The \textit{Human Mobility Model} simulates hourly movements from 8 A.M. to 10 P.M. On weekdays, individuals move from residential to work areas at time $T_d \sim U(a, b)$, stay for $T_w \sim U(c, d)$ hours, and may visit commercial areas before returning home. On weekends, they visit commercial areas at $T_e \sim U(g, h)$ with probability $P_e$, staying for $T_m \sim U(i, j)$ hours. The \textit{Disease Transmission Model} includes \textit{acquaintance} and \textit{stranger} contacts. Each individual has fixed acquaintance contacts at home ($K_r \sim U(m, n)$) and work ($K_w \sim U(o, p)$). At each timestep, infection occurs with probability $P_a$ from acquaintances and $P_s$ from strangers. The simulator parameters are calibrated using the Covid-19 $R_0$ from WHO.
 The statistics for both datasets are shown in Table~\ref{tab:dataset_summary}.
\vspace{1mm}

\begin{table}[h!]
    \centering
    \vspace{1mm}
    \caption{Experiment Result of Forecast Task. Best Performance under each setting is bolded.}\label{tab:forecast_table}
    \vspace{-2mm}
    \scriptsize
    \resizebox{0.6\linewidth}{!}{
    \begin{tabular}{lrrr|rrr}
    \toprule
    & &\multicolumn{2}{c}{UVA} &\multicolumn{2}{c}{EpiSim} \\\cmidrule{3-6}
    &PS &F1 &AUROC &F1 &AUROC \\\cmidrule[0.7pt]{1-6}
    \multirow{3}{*}{STGCN} 
    &5 &0.526 ± 0.022 &0.714 ± 0.035 &0.632 ± 0.036 &0.816 ± 0.029 \\\cmidrule{2-6}
    &10 &0.343 ± 0.028 &0.688 ± 0.010 &0.473 ± 0.025 &0.692 ± 0.043 \\\cmidrule{2-6}
    &20 &0.398 ± 0.031 &0.655 ± 0.031 &0.195 ± 0.073 &0.593 ± 0.010 \\\cmidrule[0.7pt]{1-6}
    \multirow{3}{*}{ASTGCN} 
    &5 &0.544 ± 0.038 &0.731 ± 0.060 &0.624 ± 0.030 &0.801 ± 0.005 \\\cmidrule{2-6}
    &10 &0.376 ± 0.013 &0.692 ± 0.012 &0.489 ± 0.014 &0.712 ± 0.009 \\\cmidrule{2-6}
    &20 &0.367 ± 0.009 &0.652 ± 0.011 &0.154 ± 0.045 &0.612 ± 0.025 \\\cmidrule[0.7pt]{1-6}
    \multirow{3}{*}{MSTGCN} 
    &5 &\textbf{0.721 ± 0.063} &\textbf{0.846 ± 0.013} &0.869 ± 0.084 &0.895 ± 0.010 \\\cmidrule{2-6}
    &10 &0.401 ± 0.041 &0.647 ± 0.012 &0.502 ± 0.042 &0.729 ± 0.020 \\\cmidrule{2-6}
    &20 &0.358 ± 0.024 &0.617 ± 0.065 &0.223 ± 0.035 &0.658 ± 0.056 \\\cmidrule[0.7pt]{1-6}
    \multirow{3}{*}{\textbf{EpiDHGNN}} 
    &5 &0.712 ± 0.023 &0.837 ± 0.019 &\textbf{0.918 ± 0.042} &\textbf{0.957 ± 0.065} \\\cmidrule{2-6}
    &10 &\textbf{0.576 ± 0.012} &\textbf{0.750 ± 0.008} &\textbf{0.612 ± 0.001} &\textbf{0.874 ± 0.017} \\\cmidrule{2-6}
    &20 &\textbf{0.454 ± 0.007} &\textbf{0.685 ± 0.008} &\textbf{0.298 ± 0.080} &\textbf{0.779 ± 0.071} \\\bottomrule
    \end{tabular}}
\end{table}
\textbf{Setup:} In our experiment, we utilize a 2-layer HGNN to capture neighborhood information. We perform a grid search over key hyperparameters, including hidden dimensions, learning rate, weight decay, kernal size, and $\alpha$. During training, we employ the ADAM optimizer with weight decay and gradient clipping activated to stabilize gradient updates and prevent exploding gradients. Models are trained for up to 100 epochs, with early stopping activated if the validation loss does not improve for 10 consecutive epochs. The experiments are conducted on a single NVIDIA Tesla V100 GPU with 16 GB of memory. Training time per epoch averages around 5 seconds. To enhance reproducibility, random seeds are fixed for data splitting, model initialization, and optimization processes. We run the baselines using the package EpiLearn~\cite{liu2024epilearnpythonlibrarymachine}.

\textbf{RQ1 - Performance}: Our experimental results for both source detection and forecasting are presented in Table~\ref{tab:detection_table} and Table~\ref{tab:forecast_table}, respectively, with the best performance under each setting highlighted in bold. We evaluated the models under diverse conditions to assess their robustness. For source detection, we masked timesteps of varying lengths (5, 10, and 20) to examine the models' ability to backtrack across different scenarios. Similarly, we tested forecasting performance using prediction horizons of 5, 10, and 20 timesteps. In most settings, EpiDHGNN outperforms the majority of baseline graph-based models, underscoring the advantages of hypergraph-based approaches in epidemic modeling through capturing the high-order contact interaction.

\begin{table}[!tp]\centering
\vspace{1mm}
\caption{Experiment Result of Source Detection Task. Best performance under each setting is bolded.}
\label{tab:detection_table}
\vspace{-2mm}
\scriptsize
\begin{tabular}{lrrrr|rrr}\toprule
&\multicolumn{4}{c}{UVA} &\multicolumn{3}{c}{EpiSim} \\\cmidrule{2-8}
&TSH &MRR &Hit@1 &Hit@3 &MRR &Hit@1 &Hit@3 \\\cmidrule{1-8}
\multirow{3}{*}{STGCN} 
&5 &0.491 ± 0.056 &0.300 ± 0.036 &0.650 ± 0.057 &0.242 ± 0.075 &0.145 ± 0.052 &0.282 ± 0.035 \\\cmidrule{2-8}
&10 &0.462 ± 0.064 &0.315 ± 0.074 &0.633 ± 0.024 &0.129 ± 0.039 &0.100 ± 0.103 &0.195 ± 0.052 \\\cmidrule{2-8}
&20 &0.427 ± 0.033 &0.175 ± 0.023 &0.596 ± 0.078 &0.111 ± 0.047 &0.089 ± 0.042 &0.163 ± 0.017 \\\cmidrule[0.7pt]{1-8}
\multirow{3}{*}{ASTGCN} 
&5 &0.501 ± 0.026 &0.300 ± 0.078 &0.650 ± 0.052 &0.226 ± 0.036 &0.167 ± 0.033 &0.333 ± 0.042 \\\cmidrule{2-8}
&10 &0.486 ± 0.046 &0.250 ± 0.082 &0.667 ± 0.022 &0.141 ± 0.067 &0.100 ± 0.027 &0.133 ± 0.014 \\\cmidrule{2-8}
&20 &0.416 ± 0.029 &0.205 ± 0.058 &0.650 ± 0.032 &0.118 ± 0.087 &0.076 ± 0.031 &0.100 ± 0.087 \\\cmidrule[0.7pt]{1-8}
\multirow{3}{*}{MSTGCN} 
&5 &0.618 ± 0.026 &0.417 ± 0.029 &0.767 ± 0.076 &0.333 ± 0.029 &0.167 ± 0.032 &0.400 ± 0.058 \\\cmidrule{2-8}
&10 &0.561 ± 0.026 &0.350 ± 0.050 &0.733 ± 0.058 &0.213 ± 0.058 &0.100 ± 0.026 &\textbf{0.200 ± 0.019} \\\cmidrule{2-8}
&20 &0.442 ± 0.029 &0.150 ± 0.058 &0.700 ± 0.052 &0.192 ± 0.016 &0.089 ± 0.100 &0.193 ± 0.029 \\\cmidrule[0.7pt]{1-8}
\multirow{3}{*}{\textbf{EpiDHGNN}} 
&5 &\textbf{0.704 ± 0.033} &\textbf{0.517 ± 0.076} &\textbf{0.917 ± 0.029} &\textbf{0.401 ± 0.074} &\textbf{0.200 ± 0.100} &\textbf{0.500 ± 0.100} \\\cmidrule{2-8}
&10 &\textbf{0.662 ± 0.005} &\textbf{0.500 ± 0.015} &\textbf{0.783 ± 0.029} &\textbf{0.218 ± 0.037} &\textbf{0.133 ± 0.058} &0.167 ± 0.058 \\\cmidrule{2-8}
&20 &\textbf{0.582 ± 0.031} &\textbf{0.350 ± 0.026} &\textbf{0.765 ± 0.050} &\textbf{0.219 ± 0.061} &\textbf{0.100 ± 0.100} &\textbf{0.200 ± 0.100} \\\midrule
\bottomrule
\end{tabular}
\vspace{-4mm}
\end{table}

\textbf{RQ2 - Ablation}: We conducted an ablation study on the contact pattern awareness module to investigate Question 2. As shown in Table~\ref{tab:ablation}, removing this module led to a noticeable decline in performance, indicating its crucial role in capturing individual contact patterns. The results suggest that incorporating individual contact behaviors enhances the model’s ability to encode social interactions more effectively, aligning with societal norms. This highlights the importance of modeling personalized contact dynamics in improving the overall predictive capability of our approach.

\begin{table}[!h]
\centering
\scriptsize
\vspace{1mm}
\caption{Ablation study on contact pattern awareness module.}\label{tab:ablation}
\vspace{-2mm}
\begin{tabular}{lrrrr|rrr}\toprule
& & &\multicolumn{2}{c}{UVA} &\multicolumn{2}{c}{EpiSim} \\\cmidrule{4-7}
& &Setting &MRR &Hit@1 &MRR &Hit@1 \\\cmidrule{3-7}
\multirow{6}{*}{Detection} 
&\multirow{3}{*}{w/o CT module} 
&tsh-5 &0.692 ± 0.050 &0.483 ± 0.104 &0.381 ± 0.027 &0.167 ± 0.058 \\\cmidrule{3-7}
& &tsh-10 &0.644 ± 0.032 &0.467 ± 0.058 &0.204 ± 0.020 &0.100 ± 0.013 \\\cmidrule{3-7}
& &tsh-20 &0.558 ± 0.014 &0.323 ± 0.029 &0.197 ± 0.006 &\underline{0.100 ± 0.100} \\\cmidrule{2-7}
&\multirow{3}{*}{w/ CT module} 
&tsh-5 &\textbf{0.704 ± 0.033} &\textbf{0.517 ± 0.076} &\textbf{0.401 ± 0.074} &\textbf{0.200 ± 0.100} \\\cmidrule{3-7}
& &tsh-10 &\textbf{0.662 ± 0.005} &\textbf{0.500 ± 0.005} &\textbf{0.218 ± 0.037} &\textbf{0.133 ± 0.058} \\\cmidrule{3-7}
& &tsh-20 &\textbf{0.582 ± 0.031} &\textbf{0.350 ± 0.058} &\textbf{0.219 ± 0.061} &\underline{0.100 ± 0.092} \\\cmidrule{1-7}
& & &F1 &AUROC &F1 &AUROC \\\cmidrule{4-7}
\multirow{6}{*}{Forecast} 
&\multirow{3}{*}{w/o CT module} 
&ps-5 &0.709 ± 0.004 &0.830 ± 0.003 &0.891 ± 0.003 &0.912 ± 0.007 \\\cmidrule{3-7}
& &ps-10 &0.571 ± 0.008 &0.747 ± 0.006 &0.513 ± 0.007 &0.824 ± 0.005 \\\cmidrule{3-7}
& &ps-20 &0.439 ± 0.008 &0.680 ± 0.004 &0.253 ± 0.006 &0.724 ± 0.009 \\\cmidrule{2-7}
&\multirow{3}{*}{w/ CT module} 
&ps-5 &\textbf{0.712 ± 0.023} &\textbf{0.837 ± 0.019} &\textbf{0.918 ± 0.042} &\textbf{0.957 ± 0.006} \\\cmidrule{3-7}
& &ps-10 &\textbf{0.576 ± 0.012} &\textbf{0.750 ± 0.008} &\textbf{0.612 ± 0.001} &\textbf{0.874 ± 0.004} \\\cmidrule{3-7}
& &ps-20 &\textbf{0.454 ± 0.007} &\textbf{0.685 ± 0.008} &\textbf{0.298 ± 0.080} &\textbf{0.779 ± 0.071} \\\midrule
\bottomrule
\end{tabular}
\vspace{-3mm}
\end{table}

\begin{wraptable}{r}{0.35\linewidth}
\centering
\scriptsize
\caption{Contact Pattern Prediction}\label{tab:ct_acc}
\vspace{-2mm}
\begin{tabular}{lrr|rrr}\toprule
&\multicolumn{2}{c}{UVA} &\multicolumn{2}{c}{EpiSim} \\\cmidrule{2-5}
Quantile &Range &F1 &Range &F1 \\\cmidrule{1-5}
1 &[:6] &0.795 &[:616] &0.997 \\\cmidrule{1-5}
2 &[6:11] &0.773 &[616:1788] &0.852 \\\cmidrule{1-5}
3 &[11:19] &0.809 &[1788:1847] &0.820 \\\cmidrule{1-5}
4 &[19:] &0.841 &[1847:] &0.639 \\\cmidrule{1-5}
Overall &--- &0.804 &--- &0.827 \\\midrule
\bottomrule
\end{tabular}
\vspace{-4mm}
\end{wraptable}

\textbf{RQ3 - Module Effectiveness}: To investigate whether the contact pattern is successfully captured, we evaluate the module’s performance in predicting contact existence at the location level. Specifically, locations are divided into four quantiles based on their contact intensity. For example, in the EpiSim dataset, households exhibit lower contact intensity compared to recreational locations. For each quantile, we report the prediction accuracy along with the overall accuracy in Table~\ref{tab:ct_acc}. The results suggest that the overall contact pattern is successfully reconstructed. While the UVA dataset shows little correlation between contact intensity and accuracy, the EpiSim dataset exhibits a strong negative correlation. This observation aligns with the underlying assumptions of our dataset. The UVA dataset includes hospital contacts, which may fluctuate due to patient movement, whereas in the EpiSim dataset, locations with low contact intensity likely correspond to households, where visits occur with high frequency and regularity.

\begin{figure}[h]
\hspace{-1pt}
\scriptsize
\centering
\includegraphics[width=1.0\linewidth]{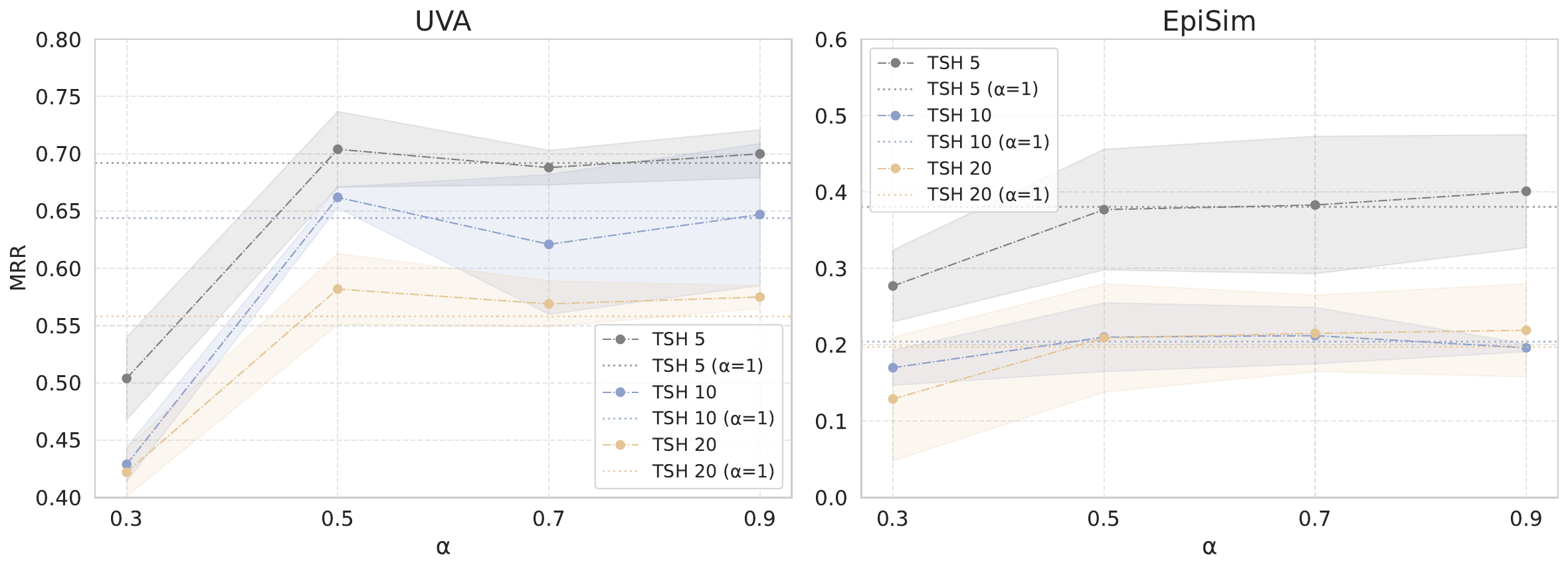}
\vspace{-3mm}
\caption{Visualization of various $\alpha$'s impact on source detection performance}
\label{fig:Alpha_analyis}
\end{figure}

We further investigate the influence of the hyperparameter $\alpha$, selecting values from {0.3, 0.5, 0.7, 0.9, 1.0}, to assess its impact on overall model performance. Lower values of $\alpha$ were not considered, as $\alpha = 0.3$ already exhibited significantly diminished performance, failing to effectively capture the model’s main task. As shown in Figure~\ref{fig:Alpha_analyis}, our results indicate that $\alpha$ has little correlation with the final performance, suggesting that it can be treated as a tunable hyperparameter for future studies.

Additionally, we observe that the model with a high timestep hidden state (TSH20) consistently outperforms other configurations when incorporating $\alpha$. This suggests that integrating contact pattern information is particularly beneficial for tasks requiring a longer temporal memory, as it helps the model better capture long-term dependencies in contact patterns. These findings highlight the importance of tuning $\alpha$ based on specific task requirements while reinforcing the advantage of incorporating contact-aware representations for long-horizon forecasting.

\begin{figure}[h]
\hspace{-1pt}
\scriptsize
\centering
\includegraphics[width=1.0\linewidth]{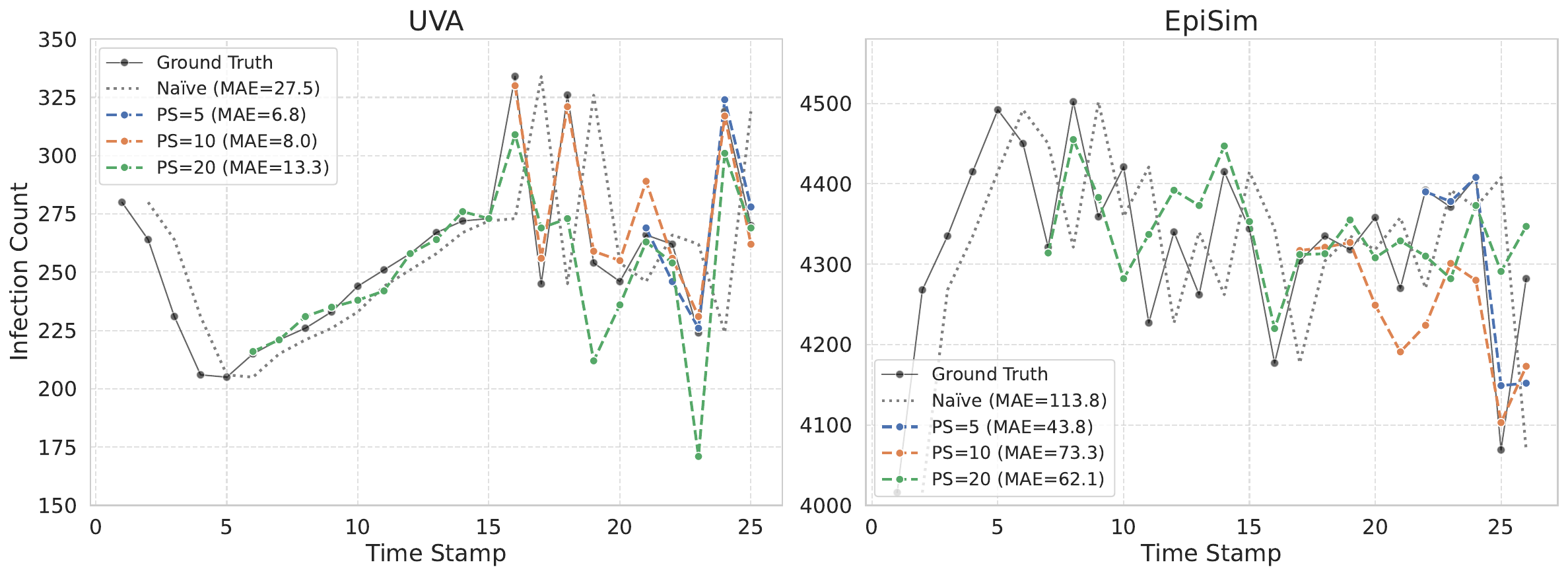}
\vspace{-3mm}
\caption{Forecast generalizability analysis. The models can successfully the future infection dynamics within various PS. We also provide the Mean Absolute Error (MAE) of Naive Model (A naive time series model forecasts future values by assuming they will be the same as the most recent observed value) and our method.}
\label{fig:forecast_exp}
\end{figure}

\textbf{RQ4 - Generalizability}: While we have demonstrated EpiDHGNN’s ability to forecast an individual’s probability of infection in RQ1, its effectiveness at capturing the population infection trend cannot be concluded. To address this, we aggregated the daily sum of infected individuals to generate population-level data across various prediction horizons with respect to time. As shown in Figure~\ref{fig:forecast_exp}, EpiDHGNN accurately captures short-term infection dynamics and effectively tracks broader fluctuations at longer time steps, albeit with reduced precision. 

%% file: section/conclusion.tex
\section{Discussion \& Conclusion}
\vspace{-3mm}
This study introduced the EpiDHGNN framework, demonstrating its ability to effectively capture the dynamics of epidemic spread through higher-order interactions in human contact networks. Through rigorous experimentation on both real-world and synthetic datasets, we validated the advantages of modeling human contact as a dynamic hypergraph, highlighting the importance of higher-order relationships and contact pattern in disease transmission.

Future improvements include exploring alternative model selections within the EpiDHGNN framework, leveraging its modular design for flexible layer substitutions. Advanced architectures in hypergraph-based learning will be investigated to enhance accuracy and efficiency. Additionally, real-world simulation remains challenging due to privacy concerns and data scarcity. Developing realistic synthetic algorithms that capture social clustering, geographic mobility, and temporal dynamics will refine simulations, improve model training, and provide a stronger benchmark for epidemic modeling.

In summary, this work underscores the power of hypergraph-based epidemic modeling and sets the stage for further exploration into both methodological advancements and data generation strategies. By enabling more accurate identification of transmission dynamics, these models hold promise for informing timely and targeted public health interventions, ultimately contributing to more effective epidemic response strategies at a broader societal level.